\newcommand{\be}{\begin{equation}}
\newcommand{\ee}{\end{equation}}

\documentstyle[epsf]{article}
\begin{document}
\onecolumn
\begin{flushright}
Alberta-Thy-9-94 \\
gr-qc/9403019
\end{flushright}
\vfill
\begin{center}
{\Large \bf Structure of the Inner Singularity of a Spherical Black Hole}\\
\vfill
 A. Bonanno\footnote[3]{Permanet address: Institute of Astronomy, University of
Catania,
Viale Andrea Doria 5, 95125 Catania, Italy}, S. Droz, W. Israel and S.M.
Morsink\\
\vspace{2cm}
Canadian Institute for Advanced Research Cosmology Program, \\
Theoretical Physics Institute\\
University of  Alberta,\\
Edmonton, Alberta, Canada T6G 2J1\\
\vspace{2cm}

\vfill
%%%%%%%%%%%%%%%%%%%%%%%%%%%%%%%%%%%%%%%%%%%%%%%%
%
%ABSTRACT
%
%%%%%%%%%%%%%%%%%%%%%%%%%%%%%%%%%%%%%%%%%%%%%%%%

\begin{abstract}
We review the evidence for and against the possibility that the inner
singularity of a black hole contains a lightlike segment which is locally
mild and characterized by mass inflation.
\end{abstract}

Submitted to: Physical Review Letters
\vfill
\end{center}
\clearpage

%%%%%%%%%%%%%%%%%%%%%%%%%%%%%%%%%%%%%%%%%%%%%%%%
%
%
%
%%%%%%%%%%%%%%%%%%%%%%%%%%%%%%%%%%%%%%%%%%%%%%%%

\twocolumn
There is currently a divergence of views concerning the structure of the
singularity inside a black hole.

It is generally agreed, and in accord with the strong cosmic censorship
hypothesis \cite{penrose}, that generically, the final singularity is probably
spacelike and of mixmaster type \cite{BLK}.

Some recent studies \cite{Hiscock,israel2} of how the internal geometry is
affected by gravitational wave-tails left in the wake of a collapse suggest the
presence, in addition, of a milder, precursory singularity which is lightlike
and
coincident with the Cauchy (inner) horizon (corresponding to infinite advanced
time $v$). It extends an infinite affine distance into the past, and tapers
to the strong final singularity at its future end (Fig. 1). The opposing view,
based on general stability arguments \cite{yurtsever} and numerical integration
of spherical models \cite{gnedin1,gnedin2}, is that this lightlike
segment of the singularity does not survive generically, but is pre-empted by
some kind of spacelike singularity.
\vspace*{0.5cm}\\
\epsfxsize=7.5cm
\epsffile{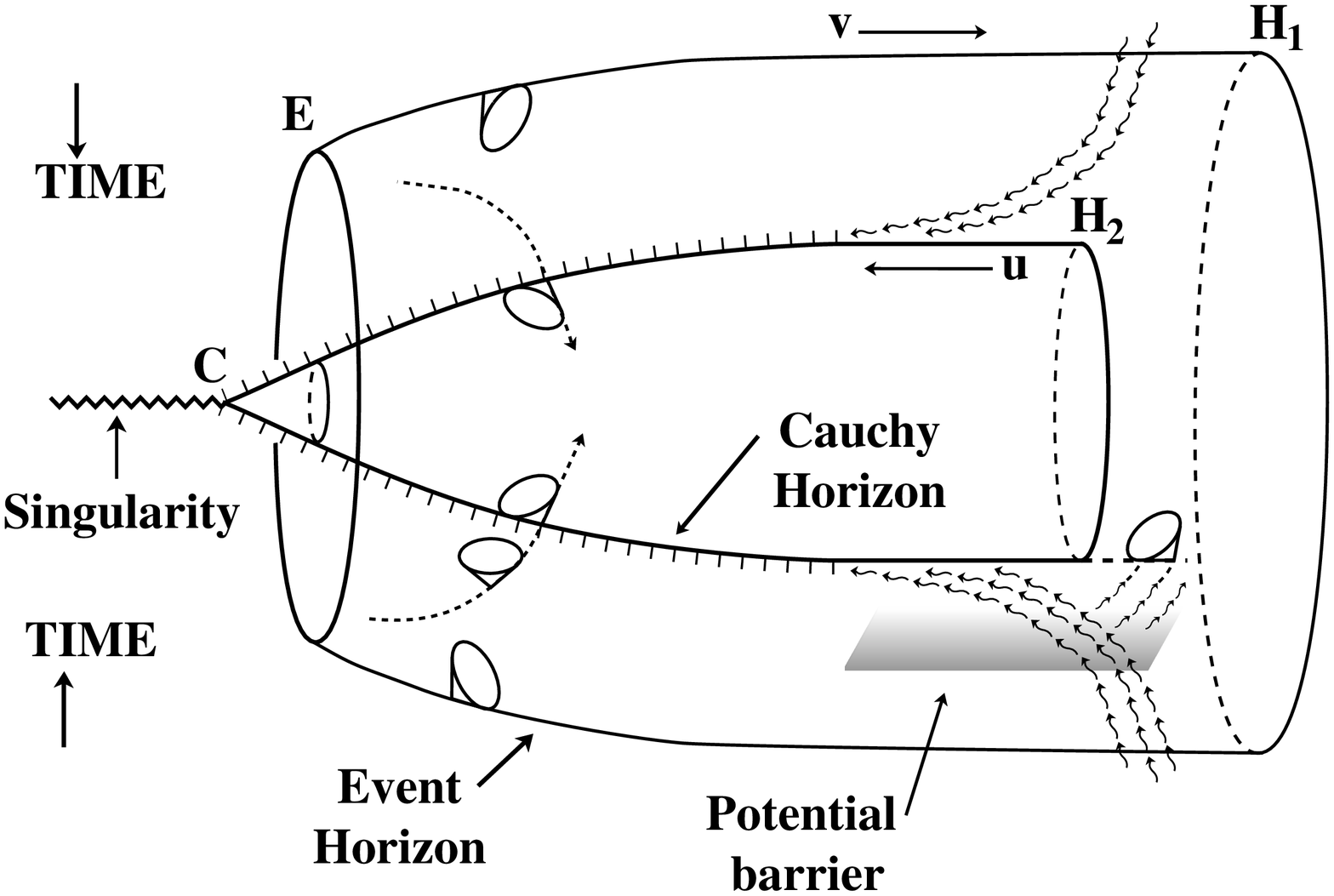}
\begin{center}
\parbox{6.5cm}{
%\baselineskip 13pt
%\vspace*{0.5cm}
\small
Fig. 1. View of the spherical black hole interior (one angular variable
suppressed), showing future light cones and a stream of infalling radiation,
partially scattered off the potential barrier, with the remainder accumulating
along the Cauchy horizon.}
\end{center}
\mbox{\hspace{1cm}}
\vspace*{0.1cm}\\
This report offers a perspective, and, hopefully, some clarification of the
issues, by recourse to analytical approximations and physical arguments.

Descent into a black hole is fundamentally a progression in time. (Recall that,
inside a spherical hole, for instance, the radial coordinate $r$ is timelike.)
To unravel the hole's internal structure is thus an {\em evolutionary} problem.
Up to the stage when curvatures begin to approach Planck levels,
the evolution can be followed even without a quantum theory of gravity.
Causality prevents our ignorance concerning the inner, high-curvature regions
from infecting the description of the outer layers afforded by well-established
(classical \cite{Hiscock,ori1,ori2,bdim} or semi-classical
\cite{balbinot,abim}) theory. (In this respect, a black hole is simpler than
a star.)

Moreover (and unlike the situation in cosmology), the initial data for the
evolution are known with precision, thanks to the no-hair property. Near the
horizon, the geometry is that of a Kerr-Newman black hole, perturbed by a tail
of gravitational waves whose flux decays as an inverse power $v^{-p}$ of
advanced time ($p=4l+4$ for a multipole of order $l$) \cite{price}.
Exploration of the hole's pre-Planckian layers thus reduces to a standard
(though
 intricate) applied-mathematical problem.

A spherical charged hole perturbed by a tail of spherisymmetric
massless scalar waves provides a simple prototype for the evolution.
Charged spherical holes have a horizon structure very similar to rotating
holes, and there is some basis \cite{ori2,bdim,israel2} for believing that the
spherical models reflect the essential qualitative features of the generic
case. (However, see Yurtsever \cite{yurtsever} for  a dissenting view.) In
what follows we restrict attention to these models.

Any spherisymmetric geometry can be described by the metric
\be
ds^2 = g_{ab} dx^a dx^b + r^2 d\Omega^2 \hspace{0.5cm}(a,b = 0,1)
\label{metric}
\ee
where $\{x^a\}$ is any pair of coordinates that label the set of 2-spheres
and $r(x^a)$ a function of these coordinates. Its gradient defines functions
$f(x^a), m(x^a)$:
\[
\left( \nabla r \right)^2 \equiv g^{ab}(\partial_a r)(\partial_b r) = f
= 1 - 2m/r  + e^2/r^2  \, .
\]
The Einstein field equations are then summed up in the two-dimensionally
covariant equations \cite{pi}
\begin{eqnarray}
\partial_a  m & = & 4 \pi r^2 T_a^{\hspace{0.5em} b} \partial_b r, \nonumber \\
r_{;ab} & = & - 4 \pi r T_{ab} + \kappa g_{ab},\label{2a}
\end{eqnarray}
which imply that the mass function $m$ satisfies the $(1+1)$-dimensional
wave equation \cite{pi}
\be
\Box m = - 16 \pi^2 r^3 T_{ab} T^{ab} + 8 \pi r f P_\bot \, . \label{wavem}
\ee
We have defined $\kappa = ( m - e^2 / r) r^{-2}, P_\bot =
T^\theta_{\hspace{0.3em}\theta}$ is the transverse pressure  and it is
understood that the stress-energy $(T_a^{\hspace{0.3em}b},P_\bot)$ does {\em
not} include Maxwellian contributions due to the hole's charge $e$ (assumed
fixed). In (\ref{2a}) and (\ref{wavem}) it has been  assumed that
$T_a^{\hspace{0.3em}a} = 0$, which holds in all cases of interest to us here.

Near the Cauchy horizon (CH), the high frequencies to which the infalling
waves get blueshifted justifies use of an ``optical`` or corpuscular
description, which models the waves as a stream of lightlike particles.
The earliest models \cite{Hiscock,pi,ori1} of black hole interiors accordingly
considered the effects of radially moving lightlike streams of dust on a
spherical charged hole.

For pure inflow (first treated by Hiscock \cite{Hiscock}) $T_{ab}$ is lightlike
and the source term of the wave equation (\ref{wavem}) vanishes:
$m$ remains bounded. It becomes a function $m(v)$ of advanced time only.
To reproduce the fallout from a radiative tail, it should take the asymptotic
form
\be
m(v) = m_0 - a v^{-(p-1)} \hspace{2em} ( v \rightarrow \infty)
\label{asympt}
\ee
where $m_0, a$ are constants. Observers falling toward
CH at radius $r_0 = m_0 -  (m_0^2 -e^2)^{\frac{1}{2}}$ and infinite $v$ see an
exponentially
blueshifted energy flux $\sim (a/r^2) v^{-p} e^{2\kappa_0 v}$
 (where $\kappa_0 = (m_0^2 -e^2)^{\frac{1}{2}} / r_0^2$ is the inner ``surface
gravity``).
But this has little effect on the geometry: the Weyl curvature scalar
\be
-\Psi_2 = (1/2) C^{\theta\varphi}_{\hspace{0.7em}\theta\varphi} =
 (m - e^2/r ) r^{-3} \label{Psi2}
\ee
remains bounded at CH.

This state of affairs changes radically when one turns on  a simultaneous
outflux \cite{pi,ori1}.
In (\ref{wavem}), $T_{ab}T^{ab}$ now functions as an exponentially divergent
source near CH. The mass function and Weyl curvature (\ref{Psi2}) diverge
like
\be
 m(v,r) \sim v^{-p} e^{\kappa_0 v} \hspace{0.5cm} (v \rightarrow \infty,
r< r_+) \label{divmass}
\ee
near CH.
This bizarre phenomenon has been dubbed ``mass inflation`` \cite{pi,helliwell}.
(It is not detectable externally: outside the event horizon EH $(r > r_+)$, $m$
stays bounded and resembles
(\ref{asympt}) at late times.)

An additional effect of the outflux is to focus the generators, and
force eventual contraction to zero radius of each lightlike 3-cylinder $v =
const.$ ,CH in particular. Until this happens to CH,  the mass inflation
singularity  remains pancake-like and locally mild \cite{ori1,bbip} in the
sense that, although tidal forces (i.e. curvatures) become infinite, they do
not grow fast enough to tear free-falling test-objects apart before they reach
CH.

However,  the analyses on which these conclusions are based contain an
assumption which is open to potentially serious criticism.
Wave tails near CH are modeled by (decoupled, separately conserved) lightlike
streams, and in the original analyses \cite{pi,ori1} it was assumed that the
outflow is turned on abruptly inside EH
($u = -\infty$, see Fig. 2) at a finite retarded time $u$.
In reality, the outflux extends back to EH and before. It is not a priori
guaranteed that its falloff at early times is rapid enough to allow the
contraction of CH to begin from an asymptotically constant radius $r_0$. If
not, the effect would be the destruction of the Cauchy horizon.

 Let us examine this further. The evolution of CH (or of any imploding
spherical light wave $V = const.$) is governed by Raychaudhuri's
equation (derivable from (\ref{2a}))
\be
d^2r/d\lambda^2  = - 4 \pi  r T_{\lambda\lambda}. \label{ray}
\ee
Here $\lambda$ is an affine parameter, and
\begin{eqnarray}
T_{\lambda\lambda} & = & T_{ab} l^a l^b, \nonumber \\  \label{TLL}
l^a & = & dx^a/d\lambda  =  - g^{ab} \partial_b V
\end{eqnarray}
is the transverse flux that focuses the lightlike generators.

Unless $T_{\lambda\lambda}$ falls off fast enough that
\be
\lambda^2 T_{\lambda\lambda} \rightarrow 0 \ \ {\rm as  }\ \ \lambda
\rightarrow - \infty,  \label{falloff}
\ee
no solution of (\ref{ray}) exists for which $r(\lambda)$ is bounded in the
remote past.
\vspace{0.5cm}\\
\epsfxsize=7.5cm
\epsffile{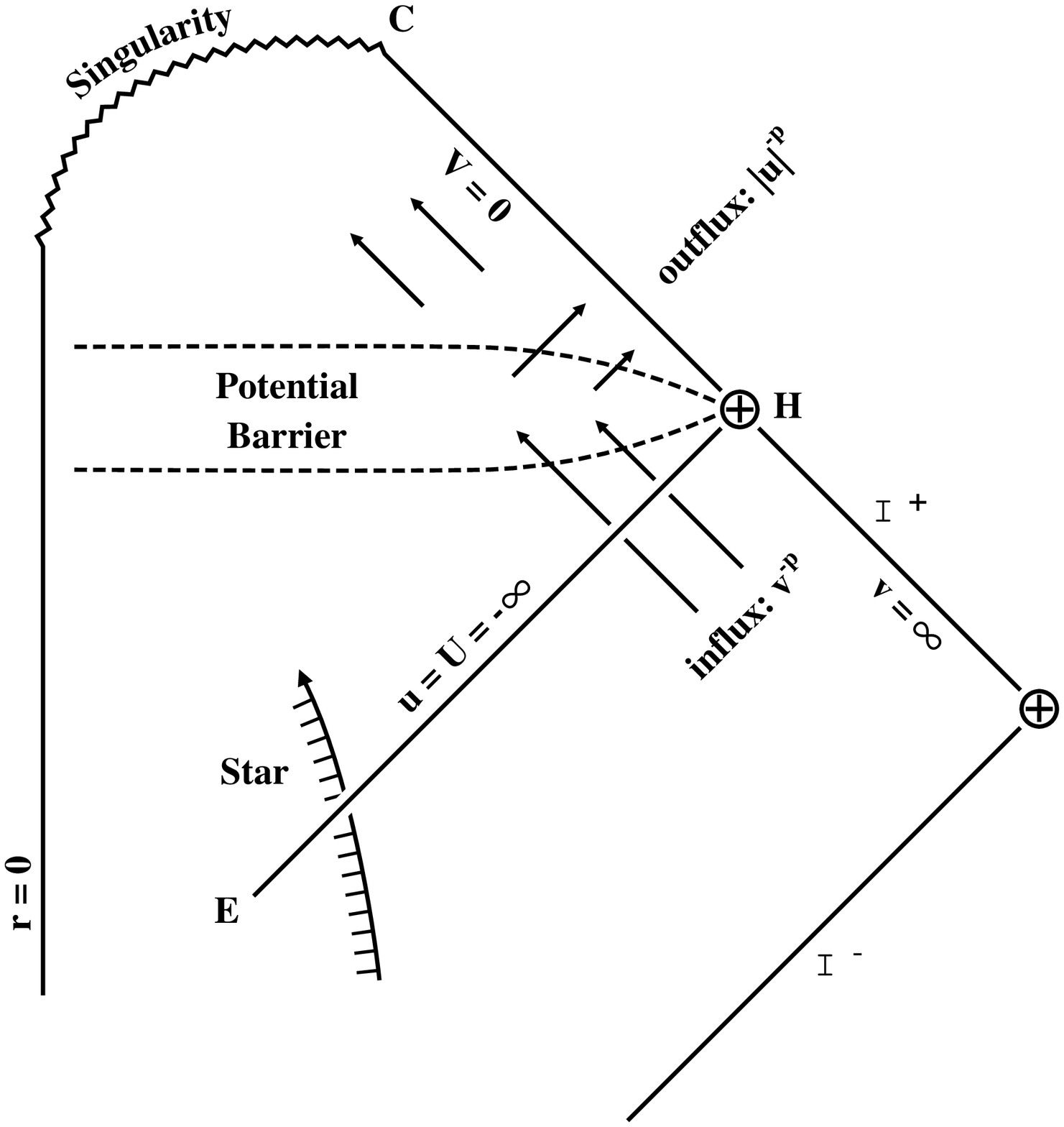}
\begin{center}
\parbox{6.5cm}{
\small
Fig. 2. Penrose conformal map of a charged spherical hole. Points H$_1$ and
H$_2$ in Fig. 1 are here telescoped into the ``corner`` H, a singular point of
this mapping.}
\end{center}
\mbox{\hspace{1cm}}
\vspace*{0.1cm}\\
We next estimate the wave flux actually incident on CH.
 This is dominated  by backscatter (approaching total reflection at late $v$)
of the infalling wave tail $T_{vv} \sim v^{-p}$ off the inner potential barrier
\cite{matzner}.
This is really a double-humped barrier-well (the potential is $2 \kappa f /
r$), which occupies a spherical shell between CH and EH concentrated near $r
\approx e^2 / m_0$ and falling off steeply near both horizons.
(Because this shell lies well above the shell of large blueshift around CH, the
geometry here at late $v$ is very close to the static
Reissner-Nordstr\"om (RN) geometry with mass $m_0$, and we may assume that the
scattering takes place on this background.) At fixed $r$
within the barrier, the infalling wave has ``time``-variation $\sim t^{-p}$,
 where $t = \frac{1}{2}(v - u)$ is the standard RN ``time`` coordinate.
Scattering generates an outflux $T_{uu} \sim \mid u \mid^{-p}$ for
$u \rightarrow -\infty$ (see Fig. 2), i.e.
\begin{eqnarray}
T_{UU} & \sim  & U^{-2} \left[ \ln{-U} \right]^{-p} \label{TUU} \\
& & ( r < e^2/m_0 , \ U \rightarrow -\infty) \nonumber
\end{eqnarray}
where we have defined Kruskal-like null coordinates regular on CH by
\[
U = - e^{-\kappa_0 u}, \hspace{1cm} V = - e^{-\kappa_0 v};
\]
$V = 0$ on CH and $U = - \infty$ along EH.

We must now compare (\ref{TUU}) with  (\ref{falloff}).
The missing link is the relation between $U$ and $\lambda$ given by
\(
dU/d\lambda  = -g^{UV},
\)
according to (\ref{TLL}). If $T^{UV}$ diverges faster than
$[\ln{-U}]^{\frac{1}{2} p}$ for $V \rightarrow 0, \ U \rightarrow - \infty$ (in
that order), then (\ref{falloff}) is violated and there is no mildly singular
lightlike segment  CH.

However, the wave equation for lightlike crossflow \cite{pi}
\[ \Box \ln{(r^{-1} g^{UV})} = - ( 3 e^2 - r^2) r^{-4}
\]
has (for $r \neq 0$ )  no diverging source term  (contrast
(\ref{wavem})) to produce a divergence of $g^{UV}$. This is evidence
that $g^{UV}$ remains bounded and that CH survives.

To lend weight to this conclusion, access to the complete solution for
the metric and the scalar field near   CH would certainly be
desirable. Approximate analytical expressions for these quantities are
actually not very difficult to construct.

In terms of lightlike coordinates $U,V$  the minimally-coupled wave equation is
\[
r \varphi_{UV} + r_U \varphi_V + r_V \varphi_U = 0
\]
for a spherisymmetric massless field $ \varphi(U,V)$.

The Einstein equations (\ref{2a}) now appear as
\begin{eqnarray}
m_U & =& - 4 \pi r^2 e^{-2 \sigma} \varphi_U^2 r_V, \nonumber \\
r_{UU} & - & 2 \sigma_U r_U  =  - 4 \pi r \varphi_U^2, \nonumber \\
(r^2)_{UV} & = &- e^{2 \sigma} \left(1 - e^2/r^2  \right), \label{sfeq}\\
\sigma_{UV} & = & (e^{2 \sigma}/r^3) (m - e^2 /r) - 4 \pi \, \varphi_U \,
\varphi_V \nonumber
\end{eqnarray}
and two further equations obtained by interchanging $U$ and $V$ in
(\ref{sfeq}).
 The subscripts indicate partial differentiation and we have written
$g_{UV} = -e^{2 \sigma}$.

We now exhibit an approximate analytical solution of these equations,
which should become increasingly reliable as one approaches the past
end of CH.

Define functions $a(U), b(V)$ by setting their derivatives
$\dot{a},\dot{b}$ equal respectively to
\nolinebreak[4]
$\varphi_U\mid_b\nolinebreak[4],~\varphi_V\mid_b$, the values on the underside
of the inner potential barrier. Define further functions
$A(U), B(V)$ by $\ddot{A} = 4 \pi r_0^2 \dot{a}^2,
\ddot{B} = 4 \pi r_0^2 \dot{b}^2$, with the boundary conditions
$A(-\infty) = B(0) = 0$. Then
\begin{eqnarray*}
\varphi & = & a(U) + b(V) \\
 & & + \ r_0^{-2} \left\{ A(U) b(V) + a(U) B(V) \right\} ,
\\
r^2 & = &r_s^2(U,V) - 2 A(U) - 2 B(V), \\
\sigma & = & \sigma_s(U,V) + r_0^{-4} A(U)B(V), \\
m  & = & m_0 + (\kappa_0^2/r_0)  \dot{A}(U) \dot{B}(V)
\end{eqnarray*}
is an approximation to the solution below the barrier.
Subscript  $s$ refers to the static RN solution (mass $m_0$,
inner-horizon  radius $r_0$) which  forms the final exterior state.
The general conditions for the  validity of the approximation,
\[
\dot{A}^2 \ll  \ddot{A}, \ \ \ \dot{B}^2 \ll  \ddot{B},
\]
are satisfied in the situation of interest to us:
\[
\left. \begin{array}{l}
A \sim [\ln{(-U)}]^{-(p-1)}, \\
 B \sim \mid -\ln{(-V)}\mid^{-(p-1)} \end{array}
 \right.
(U \rightarrow - \infty, \ V \rightarrow -0).
\]

These expressions confirm that the metric components $e^{2 \sigma}, \ r^2$
(though not their derivatives) are regular and approach the RN values
toward the past end of CH.

A more accurate global picture of the solution requires numerical integration.
Sophisticated codes for handling the spherical scalar-Einstein equations are
now available \cite{price,gomez}, and are being adapted to the charged case
by several groups \cite{gnedin1,gnedin2,carsten}.

The pioneering numerical study is due to Gnedin and Gnedin
\cite{gnedin1,gnedin2}. They consider a scalar wave pulse of finite
$v$-duration imploding into a charged hole. This initial condition was set (a)
on the event horizon in their first study \cite{gnedin1} and (b) outside
the horizon in the second one \cite{gnedin2}. Only in case (b) is an infalling
radiative tail produced (by doublescattering off the {\em external} potential
barrier).
Their results illustrate the dramatic effects of the tail. A lightlike segment
CH
is clearly evident in case (a), and much abbreviated or possibly absent in case
(b). Gnedin and Gnedin \cite{gnedin2} state that it is absent. But
it seems to us that the present numerical accuracy (see their Fig. 5) does not
warrant any firm statement. However, with the efforts now being concentrated on
this problem \cite{carsten}, a definite answer should not be long in coming.

In summary, we do not yet see any compelling reason to dismiss the naive
scenario suggested by the simplest spherical models and illustrated in Fig. 1.
This pictures a strong and ``hairy`` final singularity connected to the
asymptotically stationary and hairless outer layers of the hole by a
mildly singular lightlike ``bridge`` CH characterized by mass inflation.

This work was supported by the Canadian Institute for Advanced Research and by
NSERC of Canada. S. M. M. acknowledges the support of an Avadh Bhatia
Scholarship.

%%%%%%%%%%%%%%%%%%%%%%%%%%%%%%%%%%%%%%%%%%%%%%%%
%
%BIBLIOGRAPHY
%
%%%%%%%%%%%%%%%%%%%%%%%%%%%%%%%%%%%%%%%%%%%%%%%%
%\newpage
%\onecolumn

\end{document}